\documentclass[12pt]{iopart}

\usepackage{graphicx} 
\usepackage{xcolor}
\usepackage{amsfonts}
\usepackage{amssymb}
\usepackage{hhline}
\usepackage{booktabs}
\usepackage[linktocpage=true]{hyperref}
\usepackage[numbers,sort&compress]{natbib}
\definecolor{cadmiumgreen}{rgb}{0.0, 0.6, 0.2} 
\definecolor{ultramarine}{rgb}{0.07, 0.04, 0.56}
\definecolor{indigo(dye)}{rgb}{0.0, 0.25, 0.42}

\hypersetup{
colorlinks=true,
citecolor=ultramarine,
linkcolor=cadmiumgreen,
urlcolor=indigo(dye),
pdfauthor={},
pdftitle={},
pdfsubject={}
}
        
\begin{document}

\title[Next-decade standard sirens]{Listening Across the Cosmic Time: \\Standard Sirens from Ground- and Space-Based Missions in the Next Decade}

\author{Alberto Salvarese$^1$, Hsin-Yu Chen$^1$, Alberto Mangiagli$^2$, Nicola Tamanini$^3$}
\address{$^1$ Department of Physics, The University of Texas at Austin, 2515 Speedway, Austin, TX 78712, USA}
\address{$^2$ Max Planck Institute for Gravitational Physics (Albert Einstein Institute),
Am M¨uhlenberg 1, 14476 Potsdam, Germany}
\address{$^3$ Laboratoire des 2 Infinis - Toulouse (L2IT-IN2P3), Université de Toulouse, CNRS, F-31062 Toulouse Cedex 9, France}

\ead{ alberto.salvarese@utexas.edu}
\vspace{10pt}

\begin{abstract}
Precise measurement of the Hubble parameter will enable stringent tests of the standard model for cosmology. Standard sirens, using the luminosity distances measured by gravitational-wave observations of compact binary mergers, are expected to provide such measurements independently in the next decade. With the ground- and space-based gravitational wave observatories, the LIGO-Virgo-KAGRA (LVK) network and the Laser Interferometer Space Antenna (LISA), different types of standard sirens altogether will place constraints across a wide redshift range. In this paper, we forecast the precisions of standard siren Hubble parameter measurements and compare various scenarios, accounting for the dominant sources of systematic uncertainty. Specifically, we find a $2\%$ constraint on $H_0$, a $1.5-3\%$ constraint on $H(z)$ at $z=1$, and a $3-5\%$ constraint on $H(z)$ at $z=7$ when combining LVK and LISA standard sirens with precise redshift measurements from electromagnetic counterpart observations. We do not find a significant improvement when including standard sirens with no EM counterpart, but which rely on features in the black hole mass distribution, and the potential systematics introduced by the possible redshift evolution of such features could further degrade the measurement accuracy if not properly accounted for.

\end{abstract}

\section{Introduction} \label{sec:introduction}
The cosmic expansion rate, described by the {`Hubble parameter'} $H(z)$ \cite{Hubble_constant_first, Hubble_parameter_first}, is a fundamental quantity to measure in cosmology. High-precision measurements of $H(z)$ enable stringent tests of the standard $\Lambda$ cold dark matter ($\Lambda$CDM) model \cite{Dodelson:2003ft, Weinberg:2008zzc, Lyth:2009imm} and offer avenues for investigating alternative cosmological theories. Since the first detection of gravitational wave (GW) signals in 2015, followed by the first joint observation of a GW source and its electromagnetic (EM) counterparts in 2017 \cite{GW150914, GW170817}, GW observations have emerged as an independent tool to measure the cosmic expansion rate via the standard siren method \cite{Schutz1986}.

GW signals from compact binary coalescences (CBCs)—such as binary neutron star (BNS), binary black hole (BBH), and neutron star–black hole (NSBH) mergers—carry valuable information about the physical properties of the source systems. The amplitude of the observed GW signal can be approximated as \cite{Maggiore:2007ulw}

\begin{equation}
    h_s(f) \propto \frac{M_z^{5/6}}{2D_{\rm L}} f^{-7/6}g_s(\theta_{\rm JN}),
\end{equation}

where $s = +, \times$ denotes the GW polarization, $M_z = (1 + z)M$ is the redshifted chirp mass of the binary, $f$ is the GW frequency, $g_s(\theta_{\rm JN})$ is a function of the binary's inclination angle $\theta_{\rm JN}$, and $D_{\rm L}$ is the luminosity distance to the source. GW detectors measure a linear combination of the two polarizations:
$h(f) = F_{+}h_{+}(f) + F_{\times}h_{\times}(f)$, where $F_{+}$ and $F_{\times}$ are the detector’s antenna pattern functions \cite{Antenna_pattern_Schutz, Antenna_pattern_Forward, Maggiore:2007ulw}, which depend on the source’s sky position, polarization angle, and the detector configuration. From the amplitude of the GW signals, we can measure the luminosity distance to the GW sources. The redshifted chirp mass $M_z$ can be inferred from the GW frequency and its evolution in time. Other parameters, if not inferred independently from other measurements (such as EM observations), contribute to the measurement uncertainty of the luminosity distance \cite{Schutz1986}.

In a flat $\Lambda$CDM universe, the luminosity distance is given by

\begin{equation}
D_{\rm L} = c(1+z)\int_0^z\frac{dz'}{H(z')}, \quad \mbox{with} \quad H(z) = H_0\sqrt{\Omega_m(1+z)^3+\Omega_{\Lambda}}.
\label{eq:lumdist}
\end{equation}
Here, $H_0$ is the present-day ($z=0$) expansion rate of the universe, known as the {`Hubble constant'}, while $\Omega_m$ and $\Omega_{\Lambda}$ denote the fractional densities of matter and dark energy, respectively, with the constraint $\Omega_m + \Omega_{\Lambda} = 1$ in a flat geometry. By measuring the luminosity distance and redshift to astrophysical sources, we can measure the Hubble parameter.

However, GW observations do not directly yield the redshift of the source. Several methods have been proposed to estimate the redshift of GW sources \cite{Schutz1986, Holz_2005, Dalal_2006, Taylor_2012, DelPozzo2012, Gair_2023, Gray2020, Chen_2024}. The bright siren method utilizes EM counterparts produced by the binary, enabling redshift estimation by localizing the source and identifying its host galaxy. While this technique holds significant promise for constraining $H_0$ (e.g., to $\sim 2\%$ with $\mathcal{O}(50)$ events \cite{Chen_2018}), only one confirmed EM-bright CBC event was observed during the first three observing runs of the LIGO-Virgo-KAGRA (LVK) observatories—though other potential EM counterpart candidates have been proposed \cite{Graham_2020_counterpartcandidate, Abac_2024, Magaña_Hernandez_2024, kunnumkai2024kilonovaemissiongw230529mass}—in contrast to over $\mathcal{O}(100)$ CBCs lacking EM counterparts \cite{GWTC3catalog_2023}.

To address this limitation, alternative methods that do not require EM counterpart detection have been pursued. For example, the spectral siren method estimates redshifts by comparing the detected (redshifted) and intrinsic masses of CBC systems. The masses of CBCs can be inferred from the GW waveform; however, due to the cosmological redshift of the GW signals, the observed masses $m_z$ are related to the source-frame masses $m_s$ by $m_z = (1+z)m_s$. This method uses the distribution of observed masses and compares it with assumed or inferred source-frame mass distributions to extract redshift information \cite{Taylor_2012, Farr_2019, Ezquiaga_2022, agarwal2025blindedmockdatachallenge}. Especially, to mitigate biases from inaccurate assumptions about the mass distribution, recent approaches aim to jointly infer the intrinsic mass distribution and cosmological parameters in a self-consistent framework \cite{mastrogiovanni2021, Farah_2025, hernandez2024, Mali_2025}.

Other standard siren approaches that do not rely on EM counterparts have also been developed \cite{cousins2025stochasticsirenastrophysicalgravitationalwave, ferraiuolo2025inferringastrophysicscosmologyindividual}, including the statistical dark sirens \cite{MacLeod_2008, DelPozzo2012, Gair_2023, Fishbach_2019, Chen_2018,  Soares_Santos_2019, Gray2020, Finke_2021, Mukherjee_2021b, mancarella2022cosmologymodifiedgravitydark, Gray2022, Mukherjee_2024} and love sirens \cite{Messenger2011, Chatterjee2021, Jin_2023}. The statistical dark siren method utilizes galaxy catalogs to estimate the redshifts of all potential host galaxies within the localization volume of a GW event. By marginalizing over these candidate redshifts, one can statistically constrain cosmological parameters.
The love siren approach, applicable to systems containing at least one neutron star, exploits tidal effects imprinted on the GW signal. The nuclear matter properties of neutron stars—such as the equation of state (EoS)—are unaffected by cosmological redshift. These properties govern the tidal deformability during the inspiral phase and leave characteristic imprints on the GW waveform. If the EoS is constrained, the tidal signature can be used to estimate the source-frame mass of the neutron star. Comparing the estimated source-frame mass with the observed (redshifted) mass allows for inference of the source redshift.

In this paper, we forecast measurements of the Hubble parameter by jointly analyzing bright and spectral sirens, explicitly incorporating their dominant sources of systematic uncertainty. We do not directly consider constraints from the statistical dark siren method (however, see the discussion of `golden dark siren' below), as it remains limited by several key challenges. A primary concern is the incompleteness of galaxy catalogs \cite{Gray2020, Gray2022, Chen_2018}, particularly at higher redshifts, where the catalogs become increasingly incomplete, reducing the precision of the inference of cosmological parameters. 
An important source of systematics for dark sirens arises from the weighting scheme used to assign host probabilities to galaxies in the catalog, which is often based on observable properties such as star formation rate or color. Recent studies \cite{Hanselman_2025, Perna2025, naveed2025darkstandardsirencosmology} have demonstrated that incorrect weighting prescriptions can lead to biased estimates of cosmological parameters.
Moreover, as the analysis extends to greater distances, the localization volumes of GW events encompass a rapidly growing number of potential host galaxies, diminishing the statistical power of this method. While these limitations may be alleviated in favorable cases—namely, golden dark sirens, which are nearby and well-localized events associated with only one or a few credible host galaxies—the expected number of such events during the current LVK era is limited \cite{chen2016findingoneidentifyinghost, Nishizawa_2017, Borhanian_2020, Muttoni_2023, Chen_2024}. Golden dark sirens are anticipated to become significantly more informative with the advent of next-generation GW observatories, such as the Einstein Telescope (ET) \cite{Punturo_2010, Branchesi_2023} and Cosmic Explorer (CE) \cite{reitze2019cosmicexploreruscontribution, evans2021horizonstudycosmicexplorer}, projected to become operational beyond 2040 \cite{Chen_2024}.
Although a small number of golden dark sirens may contribute in the near term, their constraining power is expected to be modest and complementary to that of bright sirens. Accordingly, we consider their contribution to be effectively encapsulated within the LVK bright siren forecasts considered in this paper. We also exclude love sirens from our analysis due to their strong dependence on the knowledge of neutron star EoS, which remains a major open question in nuclear physics. Given the current uncertainties in the EoS, love sirens are not expected to yield strong standard siren constraints yet \cite{Haster_2020, Kashyap_2022, dhani2022cosmographybrightlovesirens}.

We focus exclusively on two specific detectors that will be available before 2040: the LIGO-Virgo-KAGRA network and the Laser-Interferometer-Space-Antenna (LISA) \cite{laserinterferometerspaceantenna}, expected to provide a full dataset around 2040. 
LISA is expected to deliver new classes of standard sirens both with \cite{Mangiagli_2022,Tamanini:2016zlh,Tamanini:2016uin,Toscani:2023gdf,Corman:2021avn,Speri:2020hwc,LISACosmologyWorkingGroup:2019mwx} and without EM counterpart association \cite{Laghi_2021,Liu:2023onj,Muttoni:2021veo}.
However there are no comprehensive studies on LISA spectral sirens in the literature so far, and for this reason for LISA we only consider bright sirens given by massive black hole binaries (MBHBs) \cite{Mangiagli_2022,Tamanini:2016zlh}. These massive binary coalescences can be detected by LISA up to $z\sim 10$ \cite{Mangiagli_2022, Auclair_2023, Laghi_2021}, and are likely to produce EM counterparts thanks to the presence of accretion disks or gas in their surroundings \cite{Armitage_2002, Milosavljević_2005, Dotti_2006, Kocsis_2006}. Other astrophysical sources—such as extreme mass ratio inspirals and stellar-origin compact binaries—are also expected to be detected by LISA \cite{Auclair_2023} and may provide supplementary cosmological information. However, they will be mostly detected at $z\lesssim 1$ and the number of events remains modest \cite{Ruiz-Rocha_2025, Buscicchio_2025}, limiting their contribution to standard sirens compared to that could be obtained with ground-based GW detectors. This is why we consider MBHBs the primary LISA standard sirens in this study.

\section{Methods}

\subsection{Bayesian framework}

To combine multiple GW events and infer $H(z)$, we adopt a Bayesian hierarchical framework. For a set of hyper-parameters $\vec{\Lambda}$ and observational data $\vec{\mathcal{D}}$, the Bayes theorem \cite{Bayes1763, jeffreys1998theory} states

\begin{equation}
    p(\vec{\Lambda}|\vec{\mathcal{D}}) = \pi(\vec{\Lambda})\frac{\mathcal{L}(\vec{\mathcal{D}}|\vec{\Lambda})}{p(\vec{\mathcal{D}})}
\end{equation}
where $\pi(\vec{\Lambda})$ represents the prior on $\vec{\Lambda}$, $\mathcal{L}(\vec{\mathcal{D}}|\vec{\Lambda})$ is the likelihood function, and $p(\vec{\mathcal{D}})$ is the evidence. For $N_{\rm obs}$ independently observed events, the likelihood of observing the set of data $\vec{\mathcal{D}}=(\vec{\mathcal{D}}^1,...\vec{\mathcal{D}}^N)$ given the hyper-parameters $\vec{\Lambda}$ can be written as \cite{Loredo_selection, mandel_2019, Vitale2020}
\begin{equation}
    \mathcal{L}(\vec{\mathcal{D}}|\vec{\Lambda}) \propto \prod_{i=1}^{N_{\rm obs}} \frac{\int d\vec{\Theta} \mathcal{L}(\vec{\mathcal{D}}^i|\vec{\Theta})p(\vec{\Theta}|\vec{\Lambda})}{\beta(\vec{\Lambda})},
    \label{eq:poisson_likelihood_rewritten}
\end{equation}
where the function 
\begin{equation}
    \beta(\vec{\Lambda}) = \int_{\vec{\mathcal{D}}>\vec{\mathcal{D}}_{\rm th}} d\vec{\Theta}d\vec{\mathcal{D}} \mathcal{L}(\vec{\mathcal{D}}|\vec{\Theta})p(\vec{\Theta}|\vec{\Lambda})
\end{equation}
quantifies the probability that a source with parameters $\vec{\Theta}$ exceeds the detection threshold $\vec{\mathcal{D}}_{\rm th}$ given certain hyper-parameters $\vec{\Lambda}$. This function accounts for selection effects arising from the fact that not all sources are equally detectable. In practice, $\beta(\vec{\Lambda})$ is typically computed by generating synthetic signals from the population model $p(\vec{\Theta}|\vec{\Lambda})$, injecting them into realistic detector noise, and processing them through the detection pipeline to determine whether they exceed $\vec{\mathcal{D}}_{\rm th}$ \cite{will_farr_selection}. The detection threshold could be a threshold on the signal-to-noise ratio produced by the source in the detector. We follow the methods described in \cite{Chen_2018} and \cite{Farah_2025} to compute the selection function for LVK bright and spectral sirens, respectively. For LISA bright sirens, we adopt the assumptions of \cite{Mangiagli_2024}.

\subsubsection{Bayesian framework: bright siren}
For the bright sirens approach, the redshifts of the GW sources are provided by EM observation of the hosts. Therefore, for $N_{\rm obs}$ independent events, we have a set of combined GW and EM data $\vec{\mathcal{D}}=(\vec{\mathcal{D}}^1_{\rm GW}, \vec{\mathcal{D}}^1_{\rm EM},...,\vec{\mathcal{D}}^{N_{\rm obs}}_{\rm GW}, \vec{\mathcal{D}}^{N_{\rm obs}}_{\rm EM})$.  We can rewrite the likelihood in Eq.~\ref{eq:poisson_likelihood_rewritten} as 

\begin{equation}
        \mathcal{L}(\vec{\mathcal{D}}|\vec{\Lambda}) = \prod_{i=1}^{N_{\rm obs}}\frac{\int d\vec{\Theta} \mathcal{L}(\vec{\mathcal{D}}^i_{\rm GW}, \vec{\mathcal{D}}^i_{\rm EM}|\vec{\Theta})p(\vec{\Theta}|\vec{\Lambda})}{\beta(\vec{\Lambda})}.
\end{equation}
We separate $\vec{\Theta}$ into relevant physical parameters ($D_{\rm L}, z$) and other physical parameters $\vec{\Theta}'$ (e.g. component masses, spins, or the binary inclination angle), and write the likelihood for the $i$-th event as

\begin{eqnarray}
\fl \int d\vec{\Theta}\mathcal{L}(\vec{\mathcal{D}}^i_{\rm GW}, \vec{\mathcal{D}}^i_{\rm EM}|\vec{\Theta})p(\vec{\Theta}|\vec{\Lambda}) 
&=& \int dD_{\rm L}dz\,d\vec{\Theta}'\,\mathcal{L}(\vec{\mathcal{D}}^i_{\rm GW}, \vec{\mathcal{D}}^i_{\rm EM}|D_{\rm L}, z,\vec{\Theta}')p(D_{\rm L}, z, \vec{\Theta}'|\vec{\Lambda}) \nonumber\\
&=& \int dD_{\rm L}dz\,d\vec{\Theta}'\,\mathcal{L}(\vec{\mathcal{D}}^i_{\rm GW}|D_{\rm L}, z,\vec{\Theta}')\mathcal{L}(\vec{\mathcal{D}}^i_{\rm EM}|D_{\rm L}, z,\vec{\Theta}') \nonumber\\
&& \quad\quad\quad\quad\quad \times p(D_{\rm L}, z, \vec{\Theta}'|\vec{\Lambda}).
\label{eq:ithlikelihood}
\end{eqnarray}
Here, we assume that the GW and EM observations are independent. We can further simplify this expression by knowing that GW data do not provide a redshift measurement in the bright sirens approach, and that EM data do not give a $D_{\rm L}$ measurement. Finally, accounting for the dependencies between the parameters in $p(D_{\rm L}, z, \vec{\Theta}'|\vec{\Lambda})$ \cite{Chen_2018, Mangiagli_2024}, Eq.~\ref{eq:ithlikelihood} becomes
\begin{eqnarray}
\fl \int d\vec{\Theta}\,\mathcal{L}(\vec{\mathcal{D}}^i_{\rm GW}, \vec{\mathcal{D}}^i_{\rm EM}|\vec{\Theta})p(\vec{\Theta}|\vec{\Lambda}) 
&=& \int dD_{\rm L}dz\,d\vec{\Theta}'\,\mathcal{L}(\vec{\mathcal{D}}^i_{\rm GW}|D_{\rm L},\vec{\Theta}')\mathcal{L}(\vec{\mathcal{D}}^i_{\rm EM}|z,\vec{\Theta}') \nonumber\\
&& \times \delta\left[D_{\rm L} - \hat{D}_{\rm L}(z,\vec{\Lambda})\right]p(z|\vec{\Lambda})p(\vec{\Theta}').
\end{eqnarray}
The Dirac $\delta$-function arises from the fact that the luminosity distance is uniquely determined once the cosmological parameters and the redshift are specified (Eq.~\ref{eq:lumdist}). $p(z|\vec{\Lambda}) = \frac{R(z)}{1+z} \frac{dV_{\rm C}}{dz}(\vec{\Lambda})$ represents the probability of having a bright CBC at redshift $z$ given the choice of hyper-parameters $\vec{\Lambda}$.
Thus, by marginalizing over the parameters
$\vec{\Theta}'$, and optionally fixing those that are tightly constrained by the presence of an EM counterpart (such as sky localization), we arrive at:
\begin{equation}
    \mathcal{L}(\vec{\mathcal{D}}|\vec{\Lambda}) = \prod_{i=1}^{N_{\rm obs}}\frac{1}{\beta(\vec{\Lambda})} \int dz \mathcal{L}(\vec{\mathcal{D}}^i_{\rm GW}|\hat{D}_{\rm L}(z, \vec{\Lambda}))\mathcal
    L(\vec{\mathcal{D}}^i_{\rm EM}|z)\frac{R(z)}{1+z}\frac{dV_{\rm C}}{dz}(\vec{\Lambda})
    \label{eq:bright_likelihood}
\end{equation}
where $\vec{\Lambda} = (H_0, \Omega_m)$.
For bright sirens, the uncertainties of redshift are relatively small. Therefore, the merger rate $R(z)$ effectively acts almost as a constant factor in the likelihood. 

\subsubsection{Bayesian framework: spectral siren}
The spectral sirens approach relies solely on information extracted from GW observations. Following the same steps outlined above, the relevant physical parameters are the detector-frame ($m_z$) and source-frame ($m_s$) component masses, $D_{\rm L}$ and $z$, and we can write the numerator of Eq.~\ref{eq:poisson_likelihood_rewritten} for the $i$-th event as:
\begin{eqnarray}
\fl \int d\vec{\Theta} \,\mathcal{L}(\vec{\mathcal{D}}^i_{\rm GW}|\vec{\Theta})p(\vec{\Theta}|\vec{\Lambda}) 
&=& \int dm_z\,dm_s\,dD_{\rm L}\,dz\,d\vec{\Theta}' \,
\mathcal{L}(\vec{\mathcal{D}}^i_{\rm GW}|m_z, m_s, D_{\rm L}, z, \vec{\Theta}') \nonumber\\
&& \times \, p(m_z, m_s, D_{\rm L}, z, \vec{\Theta}'|\vec{\Lambda}).
\end{eqnarray}

We reformulate this equation using the product rule for joint probabilities, while explicitly accounting for parameter dependencies. Additionally, we note that GW data do not directly yield redshift or source-frame mass measurements. Therefore, we have
\begin{eqnarray}
\fl &\mathcal{L}&(\vec{\mathcal{D}}^i_{\rm GW}|m_z, m_s, D_{\rm L}, z, \vec{\Theta}') 
= \mathcal{L}(\vec{\mathcal{D}}^i_{\rm GW}|m_z, D_{\rm L}, \vec{\Theta}') \nonumber\\
\fl &p&(m_z, m_s, D_{\rm L}, z, \vec{\Theta}'|\vec{\Lambda}) 
= p(m_z|m_s, z)p(D_{\rm L}|z, \vec{\Lambda})p(z|\vec{\Lambda})p(m_s|\vec{\Lambda})p(\vec{\Theta}').
\end{eqnarray}

Given the definitions of luminosity distance (Eq.~\ref{eq:lumdist}) and detector-frame mass $m_z = (1 + z) m_s$, once $z$, $\vec{\Lambda}$, and $m_s$ are specified, both $D_{\rm L}$ and $m_z$ are fully determined. Thus, the first two terms on the second line can be written as Dirac $\delta$-functions, yielding:
\begin{eqnarray}
\fl \int d\vec{\Theta}\,\mathcal{L}(\vec{\mathcal{D}}^i_{\rm GW}|\vec{\Theta})p(\vec{\Theta}|\vec{\Lambda}) 
&=& \int dm_z\,dm_s\,dD_{\rm L}\,dz\,d\vec{\Theta}'\,\mathcal{L}(\vec{\mathcal{D}}^i_{\rm GW}|m_z, D_{\rm L}, \vec{\Theta}')p(\vec{\Theta}') \nonumber\\
&& \times \delta\left[m_z - m_z(m_s,z)\right]\delta\left[D_{\rm L} - \hat{D}_{\rm L}(z, \vec{\Lambda})\right]p(z|\vec{\Lambda})p(m_s|\vec{\Lambda}).
\end{eqnarray}
The parameters that we directly measure through GW observations, and thus what actually defines the GW likelihood, are $m_z$ and $D_{\rm L}$. Therefore, we can use the property $\delta(f(x))=\frac{\delta(x-x_0)}{|f'(x_0)|}$, with $x_0$ such that $f(x_0)=0$, and rewrite the Dirac $\delta$-functions in terms of $m_s$, and $z$:
\begin{eqnarray}
\fl \int d\vec{\Theta}\,\mathcal{L}(\vec{\mathcal{D}}^i_{\rm GW}|\vec{\Theta})p(\vec{\Theta}|\vec{\Lambda}) 
&=& \int dm_z\,dm_s\,dD_{\rm L}\,dz\,d\vec{\Theta}'\,\mathcal{L}(\vec{\mathcal{D}}^i_{\rm GW}|m_z, D_{\rm L}, \vec{\Theta}') \nonumber\\
&& \times\, p(z|\vec{\Lambda})p(m_s|\vec{\Lambda})p(\vec{\Theta}') \nonumber\\
&& \times\, \delta\left[m_s - \frac{m_z}{1+z}\right]\left|\frac{\partial m_z}{\partial m_s}\right|^{-1}
\delta\left[z - \hat{z}(D_{\rm L}, \vec{\Lambda})\right]\left|\frac{\partial D_{\rm L}}{\partial z}\right|^{-1}.
\end{eqnarray}

where $\hat{z}(D_{\rm L}, \vec{\Lambda})$ represents the redshift specified by a given $D_{\rm L}$ and $\vec{\Lambda}$. 
For spectral sirens, $\vec{\Lambda}=(H_0, \Omega_m, \vec{\Lambda}_m, \vec{\Lambda}_z)$, with $\vec{\Lambda}_m$ and $\vec{\Lambda}_z$ describing the mass and redshift distributions, respectively. Accounting for the different dependencies of the parameters $\vec{\Theta}$ to the hyper-parameters, from Eq.~\ref{eq:poisson_likelihood_rewritten} we find 
\begin{eqnarray}
\mathcal{L}(\vec{\mathcal{D}}|\vec{\Lambda}) 
&=& \prod_{i=1}^{N_{\rm obs}} \frac{1}{\beta(\vec{\Lambda})} \int dD_{\rm L}\,dm_z\,\mathcal{L}(\vec{\mathcal{D}}^i_{\rm GW}|m_z, D_{\rm L}) 
\frac{R(\hat{z}, \vec{\Lambda}_z)}{1+\hat{z}} \frac{dV_{\rm C}}{d\hat{z}}(H_0, \Omega_m) \nonumber\\
&& \times\, p\left(\frac{m_z}{1+\hat{z}}\bigg|\vec{\Lambda}_m\right)
\left|\frac{\partial m_z}{\partial m_s}\right|^{-1}
\left|\frac{\partial D_{\rm L}}{\partial \hat{z}}\right|^{-1}.
\end{eqnarray}

For the inference, we adopt the framework developed in \cite{Farah_2025}, fitting for only the binary primary mass ($m_1$) distribution, which is parametrized as a combined \texttt{Mass-peak + Power-law} model \cite{Abbott_2021_pop, GWTC3population}:

\begin{eqnarray}
\fl p(m_s|\mu_m, \sigma_m, \alpha_m, m_{\rm min}, m_{\rm max}, f_p) &\propto& \left[ f_p \frac{1}{\sqrt{2\pi}\sigma_m} e^{\left(-\frac{(m_s - \mu_m)^2}{2\sigma_m^2}\right)} +\ (1 - f_p) m_s^{\alpha_m} \right]\nonumber\\
&& \times \mathcal{F}(m_{\rm min}, m_{\rm max}).
\label{eq:spectral_massdist}
\end{eqnarray}
Here, the function $\mathcal{F}$ smoothly suppresses the mass distribution near the boundaries 
$m_{\rm min}$ and $m_{\rm max}$, thus preventing sharp transitions at these mass edges. $\mu_m$, $\sigma_m$ and $f_{\rm p}$ describe the position, width and relative weight of the \texttt{Mass-peak}, while $\alpha_m$ is the \texttt{Power-law} index. As done in \cite{Farah_2025}, we checked that including  mass ratio $q=m_2/m_1$ in the inference does not alter the results.

We adopt the merger rate parametrization from \cite{Callister_2020} to model the rate evolution of BBH mergers:

\begin{equation} 
   \fl p(z|H_0, \Omega_m, \alpha_z, \beta_z, z_p) = \mathcal{C}(\alpha_{z}, \beta_{z}, z_p)\frac{1}{1+z}\frac{(1+z)^{\alpha_z}}{1+\left(\frac{1+z}{1+z_p}\right)^{\alpha_z + \beta_z}}\frac{dV_{\rm C}}{dz}(H_0, \Omega_m),
    \label{eq:spectral_redshiftdist}
\end{equation}
where $\mathcal{C}(\alpha_z, \beta_z, z_p) = 1+(1+z_p)^{-\alpha_z-\beta_z}$ is a normalization constant that ensures the distribution integrates to unity. The parameters $\alpha_z$ and $\beta_z$ are power-law indices that describe the merger rate at $z<z_p$ and $z>z_p$, respectively.

\subsection{Systematic uncertainties of standard sirens}
Constraints on cosmological parameters using combinations of different standard siren approaches have been projected (e.g.,\cite{Chen_2024, wang_2021, Yang_2021, dhani2022cosmographybrightlovesirens}). However, such projections can be overly optimistic if systematic uncertainties are not properly taken into account. In this paper, we aim to incorporate known sources of systematics and assess their impact on the precision of the resulting measurements.

\subsubsection{Systematic uncertainties: bright siren} For LVK bright sirens, known sources of systematic uncertainty have been more thoroughly investigated, and several bias mitigation methods have been developed. These systematics include GW calibration uncertainties, EM selection effects, peculiar motions of the host galaxies, and biases arising from EM constraints on the binary viewing angle \cite{Huang_2025, Sun_2020, Sun_2021, Howlett_2020, Mozzon_2022, Mukherjee_2021, Nicolaou_2020, Kumar_2025, Chen_2021_viewing_angle, Chen_2020, Muller_2024, Salvarese_2024}.

The dominant source of systematic uncertainty for LISA bright sirens arises from weak gravitational lensing due to the inhomogeneous distribution of matter along the line of sight \cite{Mangiagli_2022, Mangiagli_2024, Cusin_2021}. This effect causes magnification or de-magnification of the GW signal, thereby biasing the inferred luminosity distance. The lensing-induced uncertainty increases with redshift and becomes the primary contributor to the distance uncertainty for sources at \( z \gtrsim 0.6 \). In this work, we model the lensing effects following the approach described in \cite{Mangiagli_2024, Cusin_2021}.
More in details, we took Eq. 14 and Eq. 17 from \cite{Cusin_2021} that model the $\sigma_{D_L, lens}$ error from lensing and summed it in quadrature with the error from LISA parameter estimation $\sigma_{D_L, LISA}$  and the peculiar velocities $\sigma_{D_L, pv}$, obtained from \cite{Kocsis_2006}. The total error on luminosity distance is therefore obtained as $\sigma_{D_L} = \sqrt{\sigma^2_{D_L, LISA} + \sigma^2_{D_L, lens} + \sigma^2_{D_L, pv}}$. As LISA noise, lensing and peculiar velocities are all expected to shift the recovered luminosity distance with respect to the true value, we also shift all the original $D_L$ samples from LISA parameter estimation (centered around the injected value) by a random value extracted from a Normal distribution with zero mean and standard deviation $\sigma_{D_L}$.

\subsubsection{Systematic uncertainties: spectral siren}
For spectral sirens, the primary source of systematic uncertainty arises from the unknown source-frame mass distribution. This distribution is subject to theoretical uncertainties, including those related to stellar evolution and the history of star formation \cite{Mapelli_2020, vanSon_2022}. In recent years, parametric and non-parametric models have been developed to jointly infer the mass distribution and cosmological parameters \cite{Farr_2019, Ezquiaga_2022, Mukherjee_2022, mastrogiovanni2023icarogwpythonpackageinference, Abbott_2023_cosmic_exp, Farah_2025, hernandez2024}. Non-parametric models offer greater flexibility by making minimal assumptions about the source-frame mass distribution. However, current methods do not consider the potential redshift evolution of this distribution. The mass distribution is expected to evolve with redshift due to the stellar evolution and the evolution of the star-formation environment across cosmic time. Several studies \cite{Pierra, Ezquiaga_2022, gennari2025} have emphasized the biases that can arise when cosmological parameters are inferred without incorporating the redshift evolution in the mass distribution.

To estimate the impact of this bias, we adopt the model specified in Eq.~\ref{eq:spectral_massdist}, and allow its parameters to vary as a function of redshift. For a given hyper-parameter $\theta$ of the considered mass distribution, its evolution is modeled using a sigmoid function \cite{lalleman2025}
\begin{equation}
    \theta(z) = \theta_0+\frac{\theta_z-\theta_0}{1+\exp{ \left\{-\frac{1}{\Delta_z}(z-\bar{z}_{\theta}) \right\}}}.
    \label{eq:sigmoid_evolution}
\end{equation}
$\theta_0$, and $\theta_z$ are the values of $\theta $ at $z=0$ and $z\gg \bar{z}_{\theta}$, $\Delta_z$ controls the steepness of the evolution (the lower the steeper), and $\bar{z}_{\theta}$ represents the point at which the function transitions from lower-$z$ to higher-$z$ values. Specifically, we construct three distinct evolution scenarios (see more details in Sec. \ref{sec:LVKspectral}): (i) evolution of the \texttt{Mass-peak} parameters ($\mu_m$, $\sigma_m$), (ii) evolution of the \texttt{Power-law} parameter ($\alpha_m$), (iii) combined evolution of both \texttt{Mass-peak} and \texttt{Power-law} parameters ($\mu_m$, $\sigma_m$, $\alpha_m$). In the inference, we fit Eq.\ref{eq:sigmoid_evolution} for the possible evolution of the three parameters $\alpha_m, \mu_m,$ and $\sigma_m$ for all three scenarios, adopting the uniform priors listed in \ref{app:priors}. We then compare them to the inference assuming no evolution to estimate the bias in cosmological parameters.

\subsection{Simulations of GW and EM observations}
\subsubsection{Simulations: LVK bright sirens}
The bright sirens expected from LVK observations are dominated by BNS mergers \cite{Fragione_2021}, with possibly a small contribution from NS-BH mergers \cite{Vitale:2018wlg}. BNS mergers can produce gamma-ray bursts, kilonovae, or other types of EM counterparts \cite{Totani_2013, Wang_2016, dokuchaev2017recurrentfastradiobursts, Yamasaki_2018, pan2023repeatingfastradiobursts, Most_2023}. However, both the intrinsic BNS merger rate and the joint BNS-EM detection rate remain highly uncertain \cite{Abbott_2020, Santoliquido_2021, LVK_pop2023, VedGShah_2024}.
Additionally, a small number of golden dark sirens—well-localized events without EM counterpart—may contribute similarly to the bright siren sample. In this work, we assume a total of 50 bright sirens, with a few of them may be taken as the contributions from golden dark sirens, will be observed over the next decade \cite{LVK_pop2023, VedGShah_2024, colangeli2025brightfutureprospectscosmological}.

From a catalog of 1000 1.4-1.4 M$_{\odot}$ BNS mergers simulated assuming the expected sensitivities of LIGO-Hanford, LIGO-Livingston, Virgo, and KAGRA in the 5th observing run (O5, \texttt{
AplusDesignh.txt}, \texttt{avirgo\_O5high\_NEW.txt}, \texttt{kagra\_O5128Mpc.txt} in \cite{Abbott_2020}), we randomly select 50 events. For each event, we construct the GW likelihood $\mathcal{L}(\vec{\mathcal{D}}^i_{GW}|D_{\rm L})$ following the method developed in \cite{ChenViewingAngle}, and assign a redshift $z$ assuming a fiducial cosmology with $H_0 = 67.74$ km/s/Mpc, $\Omega_m = 0.3075$, $\Omega_\Lambda = 0.6925$, and $\Omega_k=0.0$ \cite{Planck_2015}. We refer to the corresponding Hubble parameter as $H_{\rm inj}(z)$ hereafter. We assume the redshift is perfectly measured, a reasonable assumption when comparing to the $D_{\rm L}$ uncertainty in the local Universe. 

We repeat this procedure five times to reduce the impact of statistical fluctuations, using Eq.~\ref{eq:bright_likelihood} to infer cosmological parameters at each realization, and report the mean results below.

\subsubsection{Simulations: LVK spectral sirens} 
\label{sec:LVKspectral}
To simulate a catalog of detected BBH events, we begin by generating a population of BBH with specified component masses and redshifts. We then use the publicly available \texttt{GWMockCat} package \cite{Farah_2023}, following the methods used in \cite{Farah_2025}, to apply LVK O5 selection effects and produce realistic parameter estimation posteriors.

We assume the BBH rate evolves
as Eq.~\ref{eq:spectral_redshiftdist} with parameters $\{\alpha_z, \beta_z, z_p\} = \{1.0, 3.4, 2.4\}$, as inferred in \cite{GWTC3population}. For the primary mass distribution, we adopt a \texttt{Mass-peak+Power-law} (Eq.~\ref{eq:spectral_massdist}) model where we fix $\{m_{\rm min}, m_{\rm max}, f_p\} = \{10 M_{\odot}, 78 M_{\odot}, 0.05\}$ \cite{GWTC3population}. We allow other primary mass distribution hyper-parameters to evolve in redshift following Eq. \ref{eq:sigmoid_evolution}. We consider three evolution scenarios: 
\begin{enumerate}
     \item \texttt{Mass-peak only}: $\left\{\mu_{m,0},\mu_{m,z}, \bar{z}_{\mu}, \Delta_{z,\mu},\sigma_{m,0},\sigma_{m,z}, \bar{z}_{\sigma}, \Delta_{z,\sigma}\right\} = \{30 \mbox{M}_{\odot}, 15 \mbox{M}_{\odot}, 3, 1.5,$ $ 7, 5, 3, 1.5\}$ for the evolution of the \texttt{Mass-peak} only;\\
     \item \texttt{Power-law only}: $\left\{\alpha_{m,0}, \alpha_{m,z}, \bar{z}_{m}, \Delta_{z,m}\right\}=\left\{-2.7, -3.5, 3, 1.5\right\}$ for the evolution of the \texttt{Power-law} component only;\\
     \item  \texttt{Peak+Power-law}: the combination of (i) and (ii) for the combined evolution of \texttt{Power-law} and \texttt{Mass-peak} components.
\end{enumerate}
For each scenario, the hyper-parameter values at redshift zero are based on \cite{GWTC3population}. In contrast, the hyper-parameters governing redshift evolution are not informed by specific astrophysical predictions. They are instead chosen to represent illustrative deviations from the local population, allowing us to assess the robustness of cosmological inference under a generic redshift evolution model. Since the primary goal of this study is to evaluate the potential impact of redshift evolution on spectral siren measurements, we adopt these values without asserting that they reflect the real underlying population.
The secondary mass for each BBH is assigned by drawing a mass ratio $q = m_2/m_1$ from a uniform distribution over the interval $[0, 1]$, assumed to be independent of redshift. As a result, the redshift evolution of the secondary mass follows directly from the evolution of the primary mass.

\begin{figure}[t]
    \includegraphics[width=1\linewidth]{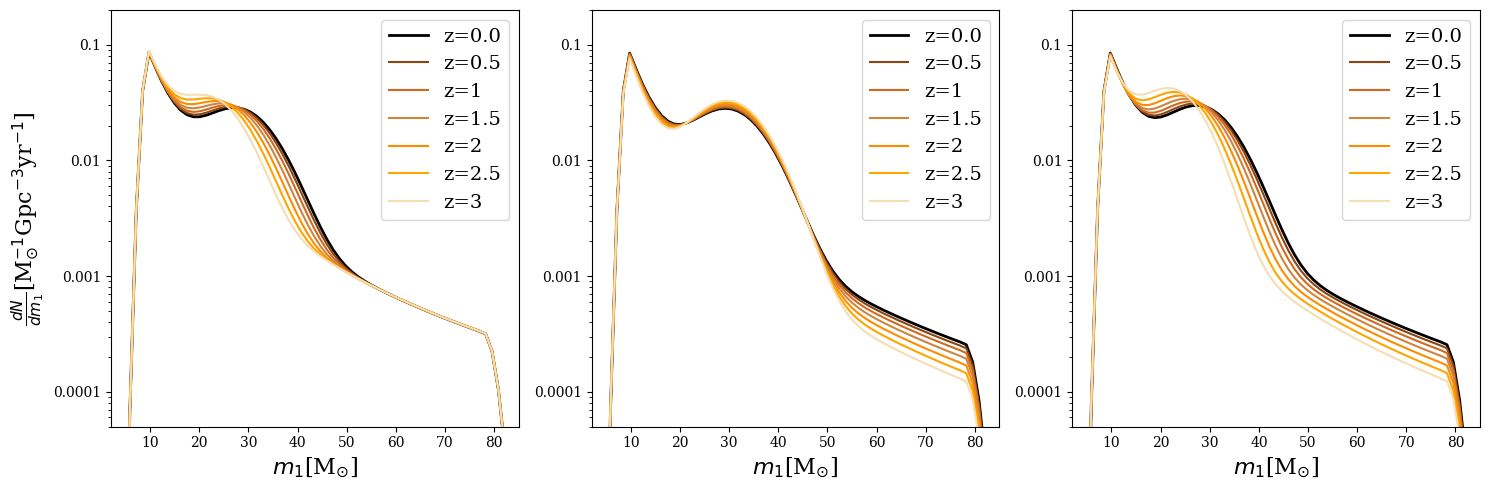}
    \caption{The three simulated primary mass distribution evolution scenarios. The left, central and right panel shows a \texttt{Mass-peak only}, a \texttt{Power-law only}, and a combined \texttt{Peak+Power-law} evolution in redshift, respectively.}
    \label{fig:evolving_mass_distributions}
\end{figure}

The three different evolution scenarios are shown in Fig.~\ref{fig:evolving_mass_distributions}, where the black line represents the BBHs mass distribution at redshift $z=0$, while the other colored lines show different z-evolution stages. The mass distribution either shifts toward lower characteristic masses (\texttt{Mass-peak only} and \texttt{Peak+Power-law}) or has a suppression of the high-mass tail (\texttt{Power-law only} and \texttt{Peak+Power-law}). For each mass evolution scenario, we simulate five independent realizations of the BBH population. Each realization includes 2800 detected events—corresponding to five year observations under LVK O5 sensitivities—which we take as a representative projection of the detection yield over the next decade.

\begin{figure}
    \centering
    \includegraphics[width=0.6\linewidth]{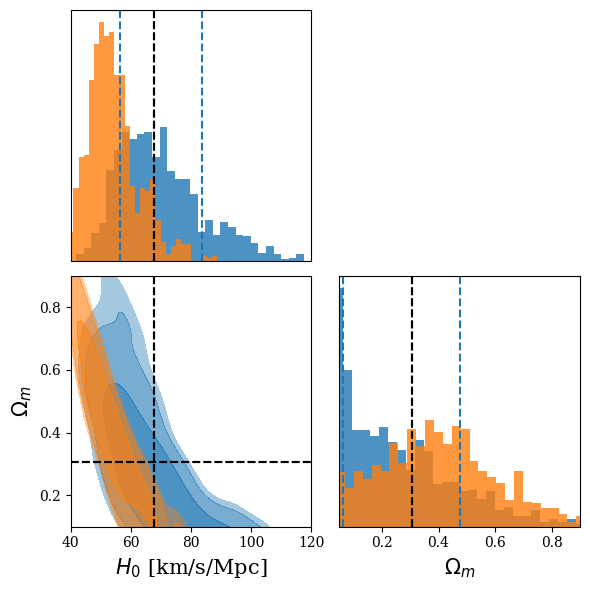}
    \caption{$H_0$ and $\Omega_m$ joint inference using a population of LVK spectral sirens evolved with the \texttt{Power-law only} evolution scenario. We show the cosmological parameter inference with (blue) and without (orange) assuming the mass distribution evolves in the inference. The different shaded regions report the $68\%$, $90\%$, and $95\%$ credible intervals. The fiducial cosmological values are shown as black dashed lines. The dashed blue lines in the marginalized posterior distributions indicate the $68\%$ credible intervals for the inference that assumes an evolving primary mass distribution.}
    \label{fig:LVK_spectral_alone_z_noz}
\end{figure}

We then infer $H(z)$ with two distinct assumptions: one assumes the primary mass distribution evolves, and another assumes it does not. Both of them adopt the simulated mass distribution model Eq. \ref{eq:spectral_massdist}.
An example of the resulting marginalized posterior distributions is shown in Fig.~\ref{fig:LVK_spectral_alone_z_noz}, where the blue (orange) contours correspond to the inference assuming an evolving (non-evolving) mass distribution under the \texttt{Power-law only} scenario. We see that while assuming an evolving mass distribution (blue) enables a proper recovery of the fiducial cosmological parameters (black dashed lines), neglecting this evolution can lead to biased inferences (orange). Similar results are found in other evolutionary scenarios.

We use the results assuming that the primary mass distribution evolves as our estimate of the statistical uncertainty. We then compare them to the ones assuming no evolution in order to estimate the systematic uncertainties originated from the redshift evolution.

\subsubsection{Simulations: LISA bright sirens}
We adopt the \texttt{\texttt{Pop3}} and \texttt{\texttt{Q3d}} MBHB formation channels described in \cite{Mangiagli_2022} as our bright CBC events (we consider the \texttt{\texttt{Q3nd}} model irrealistic as it completely neglects the time delay between the merger of two galaxies and the merger of their corresponding MBHs). 
These populations are based on the results of semi-analytical models (see \cite{Barausse12, Barausse20} and reference therein) that can track the evolution of MBHs across cosmic time. We simulate a 5-year observing period with an 80\% duty cycle.

To model the number of detectable EM counterparts, we follow the framework of \cite{Mangiagli_2022, Mangiagli_2024}, considering three scenarios: (i) \texttt{Pessimistic}: the EM emission is collimated and the obscuration given by the surrounding gas and dust is included, giving the most conservative counterpart yield; (ii) \texttt{Optimistic}: assuming no obscuration and isotropic EM emissions; and (iii) \texttt{Moderate}: an intermediate case, given by the average between (i) and (ii). The expected number of MBHB with EM counterpart is reported in Tab.~\ref{tab:scenarios_EM}. As for the previous simulations, we randomly draw the corresponding number of MBHBs for five independent realizations and infer $H(z)$ through Eq.~\ref{eq:bright_likelihood}.

For LISA bright sirens, we do not consider a perfectly measured redshift. Instead, we model the EM likelihood as a Normal distribution centered at $z = z_{\rm true} + \mathcal{N}(0,\sigma_z)$ with standard deviation $\sigma_z$ computed as described in \cite{Mangiagli_2022, Mangiagli_2024}. The value of $\sigma_z$ depends on the method and facility used to estimate the redshift of the host galaxy. For example, if the redshift is obtained with the Vera Rubin Observatory, we assume $\sigma_z = 0.031\times (1+z)$, using photometric measurements. If the redshift is determined with the Extremely Large Telescope and the galaxy is sufficiently bright, we assume that the redshift can be measured spectroscopically, leading to $\sigma_z = 10^{-3}$. If the galaxy is fainter, we still expect to be able to get an estimate of the redshift thanks to the Lyman-$\alpha$ or the Balmer break: in these cases, we assume $\sigma_z=0.2$ and $\sigma_z=0.5$, respectively. More details on the choice of $\sigma_z$ can be found in Sec. IV of \cite{Mangiagli_2022}.

\begin{table}
\begin{center}
    \begin{tabular}{l c c c}
        \toprule
        & \texttt{Pessimistic} & \texttt{Moderate} & \texttt{Optimistic} \\
        \midrule
        \texttt{Pop3} & 2 & 4 & 6 \\
        \texttt{Q3d}    & 3 & 9 & 15 \\
        \bottomrule
    \end{tabular}
    \caption{Number of LISA bright sirens over the 5-year observing period for different MBHB formation channels and EM counterpart detection scenarios \cite{Mangiagli_2024}. }
    \label{tab:scenarios_EM}
\end{center}
\end{table}

\subsection{Combination of sirens}
\begin{figure}
    \centering
    \includegraphics[width=0.5\linewidth]{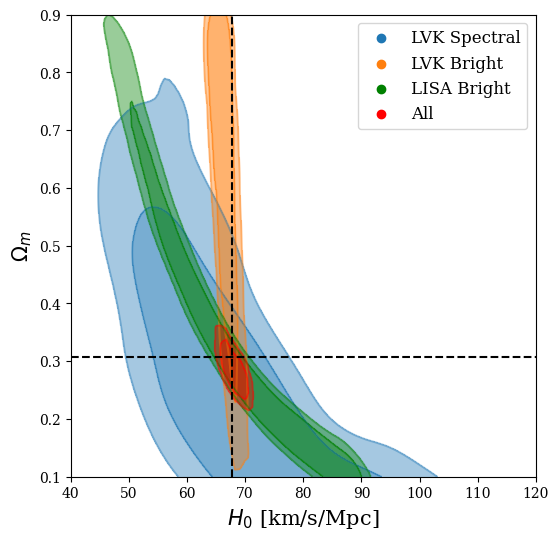}
    \caption{An example of the inferred cosmological parameter posterior from different sirens and the combined posterior. The blue, orange, and green contours show the joint ($H_0$, $\Omega_m$) posterior inferred from the LVK spectral (assuming the mass distribution evolves following the \texttt{Power-law only} scenario), LVK bright, and \texttt{Pessimistic} \texttt{Pop3} LISA bright sirens, respectively. The red contour shows the combined posterior of the three. The black dashed lines denote the injected $H_0$ and $\Omega_m$ values. The different shaded regions represent the 68\% (dark) and 90\% (light) credible intervals.}
    \label{fig:combined_posteriors}
\end{figure}
Both LISA bright sirens and LVK spectral sirens provide constraints on $H(z)$ at higher redshifts \cite{Ezquiaga_2022, Farah_2025, Mangiagli_2024}, while LVK bright sirens are particularly well-suited for measuring the local expansion rate, $H_0$. Different types of sirens provide complementary information, allowing for constraints on the cosmic expansion rate over a broad range of redshifts. To combine the information from different siren types, we adopt a sequential inference strategy: the posterior distribution obtained from one class of sirens is used as the prior for the inference with another class. This procedure is mathematically equivalent to a single, joint inference over all the considered datasets, provided that the datasets are statistically independent and the same prior is used. We adopt this approach for computational efficiency, and because it enables direct access to the inferred posteriors from individual classes of sirens.

An example of the resulting posteriors from the individual methods and their combination is given in Fig.~\ref{fig:combined_posteriors}, which shows how combining standard siren methods can improve cosmological constraints in the ($H_0$, $\Omega_m$) parameter space. 
In particular, LISA bright and LVK spectral sirens are both sensitive to $H(z)$ at $z > 0$ and exhibit an anti-correlation between $H_0$ and $\Omega_m$. This arises from the definition of $D_{\rm L}$ (Eq. \ref{eq:lumdist}) in a flat $\Lambda$CDM universe: for a given luminosity distance measurement, an increase in $H_0$ can be compensated by a decrease in $\Omega_m$, and vice versa. This degeneracy is more pronounced at higher redshifts, where $D_{\rm L}$ becomes sensitive to both parameters. In contrast, LVK bright sirens, which are predominantly located at low redshift, provide direct and precise constraints on $H_0$, effectively breaking the degeneracy and improving the overall precision of the cosmological inference.

\section{Results}

\begin{figure}
    \centering
    \includegraphics[width=1\linewidth]{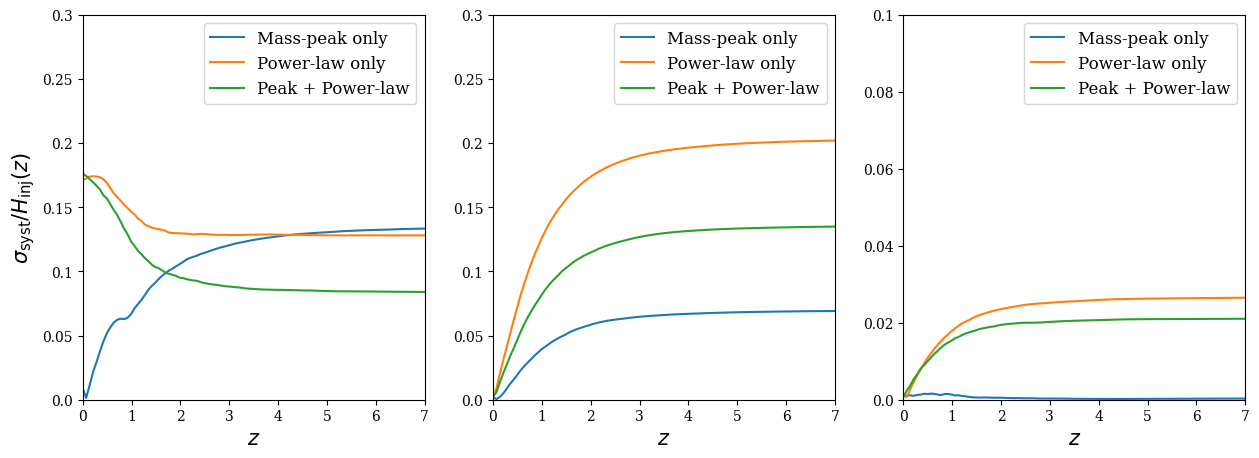}
    \caption{Relative systematic uncertainties for LVK spectral sirens. We estimate the systematics by comparing the absolute difference between the medians of the $H(z)$ posteriors inferred assuming evolving and non-evolving primary mass distribution, for three different evolution scenarios: \texttt{Mass-peak only}, \texttt{Power-law only}, and \texttt{Peak+Power-law}. Left panel: LVK spectral sirens only. Central panel: combination of LVK spectral and LVK bright sirens. Right panel: combination of LVK spectral, LVK bright, and LISA bright sirens (\texttt{Pop3 Pessimistic} scenario). Combining all classes of sources reduces the relative systematic uncertainties across redshift and mitigates the variability between different evolution scenarios.}
    \label{fig:noz_SS_systematics}
\end{figure} 

In Fig.~\ref{fig:noz_SS_systematics}, we show the relative systematic uncertainty in $H(z)$ using the LVK spectral sirens. For each simulated realization described in Sec. \ref{sec:LVKspectral}, we take the $H_0$ and $\Omega_m$ posterior samples and compute $H(z)$ following Eq.~\ref{eq:lumdist}. We compute the absolute difference $\sigma_{\rm syst}$ between the medians of the $H(z)$ posterior samples inferred assuming evolving and non-evolving primary mass distribution. We then report the average of this quantity over five independent realizations. 

In the case where only LVK spectral sirens are considered (left panel), we find that different mass evolution scenarios yield systematic biases ranging from $\sim6\%$ (\texttt{Mass-peak only}) to $\sim14\%$ (\texttt{Power-law only}) at redshift $z=1$, and from $\sim8\%$ (\texttt{Peak+Power-law}) to $\sim13\%$ (\texttt{Mass-peak only}) at $z=7$. 
By combining LVK spectral and LVK bright sirens (central panel), the systematic uncertainty at $z=0$ is negligible due to the strong constraint on $H_0$ provided by the bright sirens. However, the systematic increases with redshift: from $\sim 4\%$ (\texttt{Mass-peak only}) to $\sim 12\%$ (\texttt{Power-law only}) at $z=1$, and from $\sim 7 \%$ (\texttt{Mass-peak only}) to $\sim 20\%$ (\texttt{Power-law only}) at $z=7$. This behavior arises from the anti-correlation between $H_0$ and $\Omega_m$: fixing $H_0$ around its injected value constrains $\Omega_m$ to values that may be biased when the underlying mass distribution is incorrectly modeled. As a result, even if $H_0$ is correctly recovered, the inferred $H(z)$ deviates from the injected expansion history at $z>0$. 
When all methods are combined (right panel), the joint constraint on $(H_0, \Omega_m)$ becomes significantly tighter, effectively suppressing the impact of an incorrect modeling of the mass distribution. The right panel refers to the \texttt{Pop3 Pessimistic} scenario for LISA bright sirens; however, we find that all other EM counterpart scenarios yield similar results.

We emphasize that the systematic shown in Fig.~\ref{fig:noz_SS_systematics} depends on the specific redshift evolution assumed. Varying the rate at which a particular mass distribution feature evolves with redshift can significantly alter the resulting systematic bias. 

\begin{figure}[t]
    \centering
    \includegraphics[width=1\linewidth]{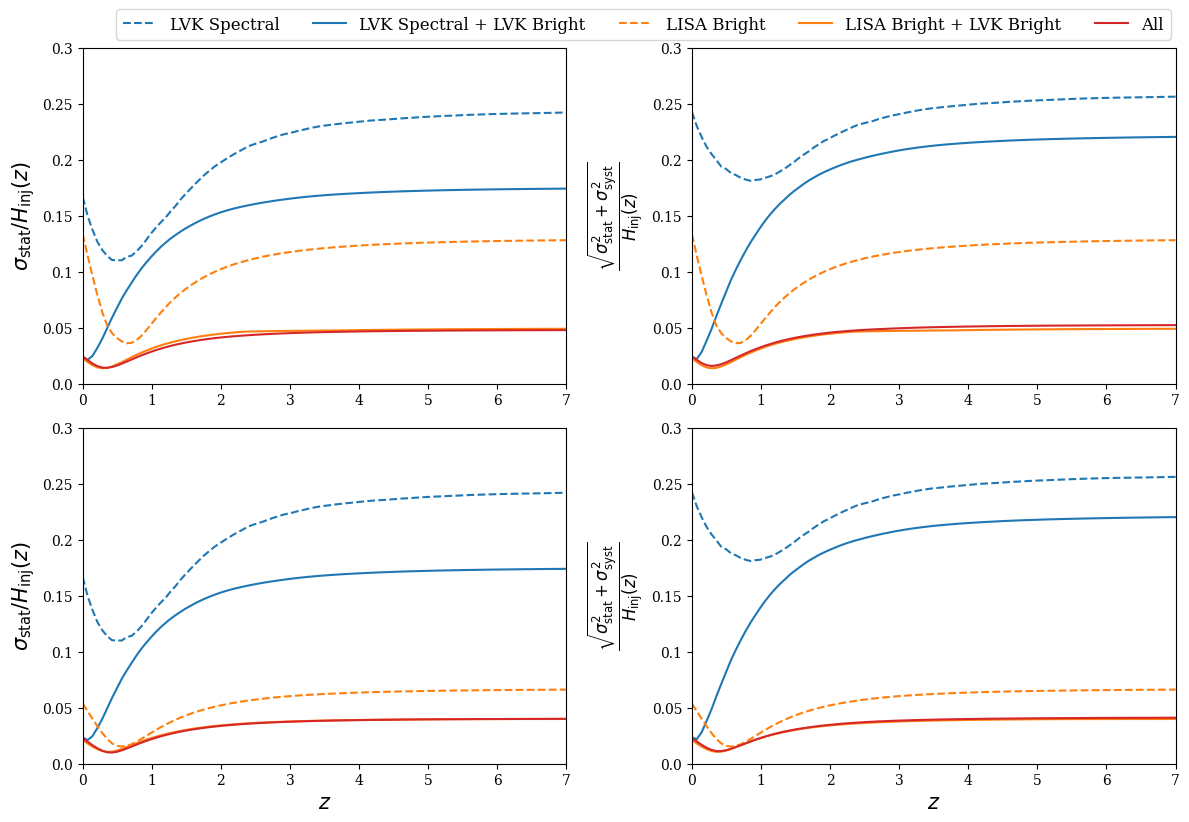}
    \caption{Left panels: Relative statistical uncertainties on $H(z)$ obtained from the different methods considered, shown for the LVK bright sirens, LVK spectral sirens with a \texttt{Peak+Power-law} evolving primary mass distribution, and the LISA bright siren \texttt{Pop3} formation channel. The upper and lower panels correspond to the LISA bright siren \texttt{Pessimistic} and \texttt{Optimistic} EM counterpart scenarios, respectively. Right panels: same as the left panel but including the spectral siren systematic ($\sigma_{\rm syst}$) uncertainties.
    } 

    \label{fig:results_Pop3}
\end{figure}

\begin{figure}[t]
    \centering
    \includegraphics[width=1\linewidth]{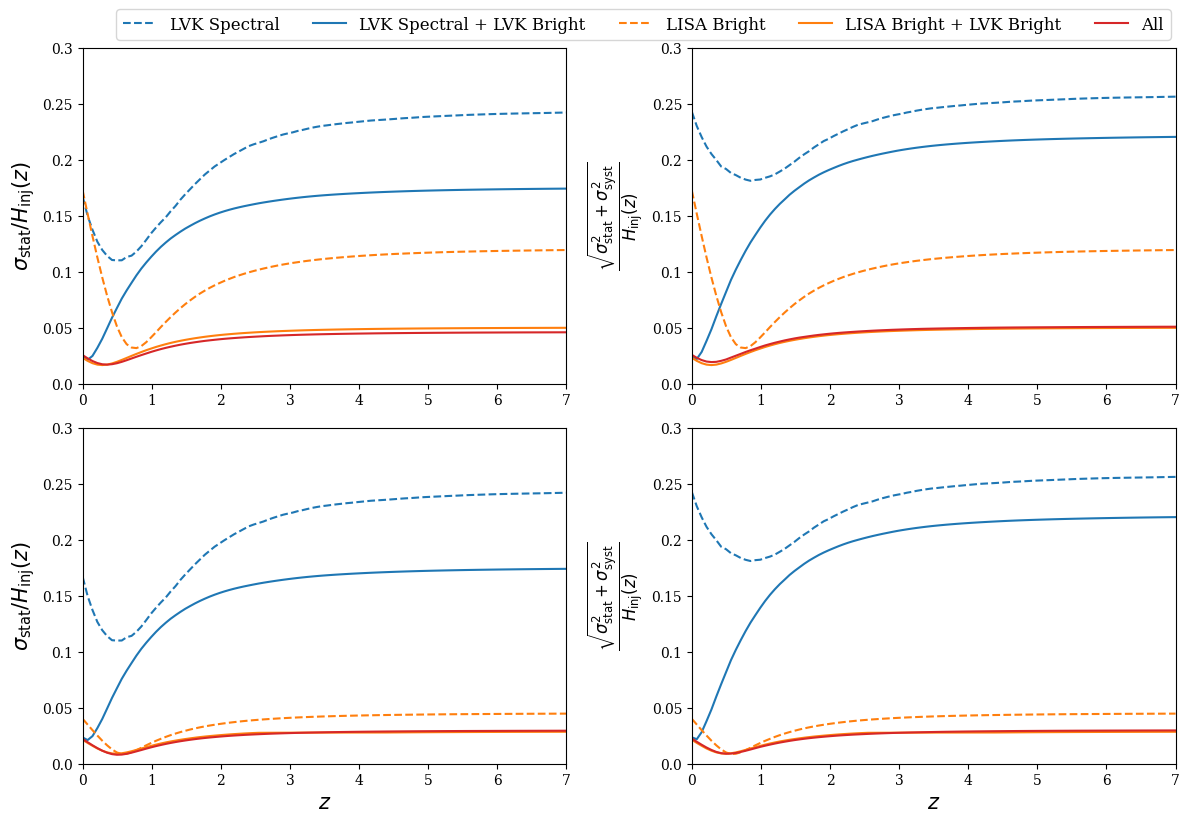}
    \caption{Same as Fig.~\ref{fig:results_Pop3}, but for the LISA bright siren \texttt{Q3d} MBHB formation channel.}
    \label{fig:results_Q3d}
\end{figure}

In Figs.~\ref{fig:results_Pop3} (\texttt{Pop3} formation channel) and~\ref{fig:results_Q3d} (\texttt{Q3d} formation channel), we compare the $H(z)$ uncertainties using different standard sirens and their combinations. The upper and lower panels correspond to the LISA bright siren \texttt{Pessimistic} and \texttt{Optimistic} EM counterpart scenarios, respectively. The left panels present the relative $1\sigma$ statistical uncertainties on $H(z)$, $\sigma_{\rm stat}$, for LVK spectral, LVK bright, and LISA bright sirens individually, as well as for their combined results.  We compute $\sigma_{\rm stat}$ as half the width of the symmetric 68\% credible interval of $H(z)$ at each redshift $z$. The right panels add in the systematic uncertainties estimated for the LVK spectral sirens, as given in Fig.~\ref{fig:noz_SS_systematics}. Here we take the \texttt{Peak+Power-law} mass evolution scenario for the LVK spectral sirens, as it represents the smallest systematic at high redshifts among the considered evolution scenarios. We note that the statistical uncertainties associated with the other two mass evolution scenarios are similar to that of the \texttt{Peak+Power-law} evolution. 

We find that:

\begin{itemize}
    \item \emph{With only one type of sirens}: The LVK spectral sirens method alone (blue dashed line), when accounting for the redshift evolution of the source-frame primary mass distribution, achieves statistical uncertainties below $\sim25\%$ across the entire redshift range. In particular, the inferred $H(z)$ precision reaches $\sim 14\%$ at $z=1$ and $\sim 25\%$ at $z=7$. LISA bright sirens alone (orange dashed line) constrain $H(z)$ to $\sim5\%$ at $z=1$ and $\sim13\%$ at $z=7$ in the \texttt{Pessimistic} scenario for both \texttt{Pop3} and \texttt{Q3d}. In the \texttt{Optimistic} scenario, the two formation channels give slightly different results, due to the larger difference in the number of EM counterparts. 
    At $z=1$ ($z=7$), the uncertainty reduces to $\sim 3\%$ ($\sim 7\%$) for \texttt{Pop3}, and to $\sim 2\%$ ($\sim 5\%$) for \texttt{Q3d}.

    \item \emph{With two types of sirens}: As expected, the LVK bright sirens significantly improves the $H(z)$ constraints at low redshifts. The statistical uncertainty on $H_0$ is $\sim 2\%$. Due to the anti-correlation between $H_0$ and $\Omega_m$ (see Fig. \ref{fig:combined_posteriors}), the constraints on $H_0$ improves the precision of $H(z)$ at higher redshifts.
    We find that the statistical uncertainty on $H(z)$ remains below $20\%$ across all redshifts when combining LVK spectral and LVK bright sirens, and below $5\%$ when combining LISA bright and LVK bright sirens. In particular, the tightest constraint is given by the \texttt{Q3d} \texttt{Optimistic} scenario (15 LISA bright sirens), which achieves a $H(z)$ precision of $\sim 1.5\%$ at $z=1$ and $\sim 3\%$ at $z=7$.

    \item \emph{With all three types of sirens}: Adding the LVK spectral sirens to the LISA bright + LVK bright sirens only leads to marginal improvements \emph{even} in the LISA bright siren \texttt{Pessimistic} scenario. In the LISA bright siren \texttt{Optimistic} scenario, the contribution from LVK spectral sirens becomes negligible, as the $H(z)$ constraints are already dominated by LISA bright sirens.

    \item \emph{With the spectral siren systematics}: Even if we consider the smallest systematic at high redshifts using the \texttt{Peak+Power-law} mass evolution scenario (green line in the left panel of Fig.~\ref{fig:noz_SS_systematics}), the systematic uncertainty dominates over the statistical uncertainty in all cases involving the LVK spectral sirens. Consequently, if the systematic in the LVK spectral sirens are not accounted for, the most precise constraints on $H(z)$ come from the combination of LISA and LVK bright sirens.
\end{itemize}

\section{Conclusions}

In this paper, we estimate the standard siren measurements of Hubble parameter within the next decade, considering their dominant systematics. 

We find that even in the \texttt{Pessimistic} EM counterpart scenario, LISA bright sirens offer tighter constraints on $H(z)$ than LVK spectral sirens. The difference is mainly due to the sensitivity of LISA to higher redshift binaries, the precise determination of the luminosity distance, and the precise determination of the redshifts for bright sirens. On the other hand, the spectral sirens are less precise due to the uncertainty of mass distribution and its evolution. 

When combining LISA and LVK bright sirens, we find $\sim 2\%$ constraint on $H_0$, $1.5-3\%$ constraint on $H(z)$ at $z=1$, and $3-5\%$ constraint on $H(z)$ at $z=7$. We do not find significant improvement when adding in the LVK spectral sirens, and the potential systematics introduced by them could further degrade the measurements.

However, spectral sirens can contribute substantially with next-generation ground-based GW interferometers, which are expected to provide sub-percent $H(z)$ measurements \cite{Chen_2024, Ezquiaga_2022} due to their higher detection rates and redshift reach. Therefore, mitigating the systematic associated with the mass evolution of the binary population will be essential to achieve such precision. 
This could be achieved through a combination of improved population-synthesis modeling, inference of population properties from large GW datasets with next-generation detectors, and joint constraints from independent probes \cite{Ezquiaga_2022, vanSon_2022, Muttoni_2023, Mastrogiovanni_2023, Ye_2024}. In particular, future bright siren measurements could be used to calibrate the primary mass distribution for spectral sirens, including its possible redshift evolution, exploiting the fact that a consistent mass model should yield the same cosmological parameters. Such cross-calibration could reduce modeling uncertainties and improve the robustness of cosmological inference from spectral sirens.

\section*{Data availability statement}
The data that support the findings of this study are available upon reasonable request from the
authors.

\section*{Acknowledgment}
The authors would like to thank Gabriele Perna for the LIGO Scientific Collaboration internal review. A.S. and H.-Y. C. are supported by the National Science Foundation under Grant PHY-2308752 and Department of Energy Grant DE-SC0025296. The authors are grateful for computational resources provided by the LIGO Laboratory and supported by National Science Foundation grants PHY-0757058 and PHY-0823459. This is LIGO Document Number P2500382.
N.T.~acknowledges support form the French space agency CNES in the framework of LISA, and from the Agence Nationale de la Recherche (ANR) through the MRSEI project ANR-24-MRS1-0009-01.

\appendix
\section{Inference priors}
\label{app:priors}
We report the uniform prior ranges adopted in our inferences in Table~\ref{tab:priors}. For $\bar{z}$, we choose the range $[1, 10]$, as current observational data disfavor significant evolution of the black hole mass distribution below redshift $z \sim 1$ \cite{lalleman2025}.

\begin{table*}[h]
\centering
\setlength{\tabcolsep}{10pt}
\renewcommand{\arraystretch}{1.3}
\scalebox{0.85}{
    \begin{tabular}{ll}
    \textbf{Cosmological parameters}\\
    \begin{tabular}{cc}
    $H_0$ & $\Omega_m$ \\
    $(30, 120)$ & $(0.01, 0.99)$
    \end{tabular} \\[5mm]
    
    \textbf{Evolving mass distribution parameters}\\
    \begin{tabular}{cc}
    $m_{\rm min}$ & $m_{\rm max}$ \\
    $(1.0, 30.0)$ & $(30.0, 100.0)$
    \end{tabular} \\[3mm]
    
    \begin{tabular}{cccc}
    $\alpha_{m 0}$ & $\alpha_{mz}$ & $\bar{z}_{\alpha_m}$ & $\Delta_z$ \\
    $(-5.0, -1.0)$ & $(-5.0, -1.0)$ & $(1.0, 10.0)$ & $(0.0, 5.0)$ \\
    \end{tabular} \\[3mm]
    
    \begin{tabular}{cccc}
    $\mu_{m 0}$ & $\mu_{mz}$ & $\bar{z}_{\mu_m}$ & $\Delta_z$ \\
    $(1.0, 60.0)$ & $(1.0, 60.0)$ & $(1.0, 10.0)$ & $(0.0, 5.0)$ \\
    \end{tabular} \\[3mm]
    
    \begin{tabular}{cccc}
    $\sigma_{m 0}$ & $\sigma_{mz}$ & $\bar{z}_{\sigma_m}$ & $\Delta_z$ \\
    $(1.0, 15.0)$ & $(1.0, 15.0)$ & $(1.0, 10.0)$ & $(0.0, 5.0)$
    \end{tabular}\\[5mm]
    \textbf{BBH redshift distribution parameters}\\
    \begin{tabular}{ccc}
    $\alpha_z$ & $\beta_z$ & $z_p$\\
    $(0.0, 5.0)$ & $(0.0, 5.0)$ & (0.0, 8.0) \\
    \end{tabular}
    \end{tabular}
}
\caption{Prior ranges for cosmological and population hyper-parameters used in the hierarchical inference.}
\label{tab:priors}
\end{table*}

\bibliographystyle{iopart-num}
\bibliography{ref}
\end{document}